\documentclass[letterpaper, 10pt, conference]{ieeeconf}
\IEEEoverridecommandlockouts
\usepackage{amsmath,amsfonts}
\usepackage{algorithmic}
\usepackage{algorithm}
\usepackage{array}
\usepackage[caption=false,font=normalsize,labelfont=sf,textfont=sf]{subfig}
\usepackage{textcomp}
\usepackage{stfloats}
\usepackage{url}
\usepackage{verbatim}
\usepackage{graphicx}
\usepackage{cite}
\usepackage{gensymb}
\usepackage{xcolor}
\usepackage{graphicx}
\usepackage{textcomp}
\usepackage{gensymb}
\usepackage{tabularx}
\usepackage{amssymb}
\usepackage{wrapfig}
\usepackage{float}
\usepackage{multicol}
\usepackage{xcolor}
\usepackage{siunitx}
\usepackage{bm}
\usepackage{changepage}
\usepackage{setspace}
\usepackage{color,soul}

\DeclareMathOperator*{\argmin}{arg\,min}

\title{Design and Development of a Lorentz Force-Based MRI-Driven Neuroendoscope}

\author{Martin Francis Phelan III$^{1,2}$, Nihal Olcay Dogan$^{1,3}$, Jelena Lazovic$^{1}$, Metin Sitti$^{1,2,3,4}$%
\thanks{$^{1}$Physical Intelligence Department, Max Planck Institute for Intelligent Systems, Stuttgart 70569, Germany  {\tt\small phelan@is.mpg.de}}%
\thanks{$^{2}$Department of Mechanical Engineering, Carnegie Mellon University, Pittsburgh, PA 15213, USA}%
\thanks{$^{3}$Institute for Biomedical Engineering, ETH Zurich, Zurich 8092, Switzerland}%
\thanks{$^{4}$College of Engineering and School of Medicine, Ko\c{c} University, Istanbul 34450, Turkey}%
}

\begin{document}


\maketitle
\thispagestyle{empty}
\pagestyle{empty}

\begin{abstract}
The introduction of neuroendoscopy, microneurosurgery, neuronavigation, and intraoperative imaging for surgical operations has made significant improvements over other traditionally invasive surgical techniques. The integration of magnetic resonance imaging (MRI)-driven surgical devices with intraoperative imaging and endoscopy can enable further advancements in surgical treatments and outcomes. This work proposes the design and development of an MRI-driven endoscope leveraging the high (3-7 T), external magnetic field of an MR scanner for heat-mitigated steering within the ventricular system of the brain. It also demonstrates the effectiveness of a Lorentz force-based grasper for diseased tissue manipulation and ablation. Feasibility studies show the neuroendoscope can be steered precisely within the lateral ventricle to locate a tumor using both MRI and endoscopic guidance. Results also indicate grasping forces as high as 31 mN are possible and power inputs as low as 0.69 mW can cause cancerous tissue ablation. These findings enable further developments of steerable devices using MR imaging integrated with endoscopic guidance for improved outcomes.
\end{abstract}

\section{Introduction}
\begin{figure}[t]
\centerline{\includegraphics[scale=0.25]{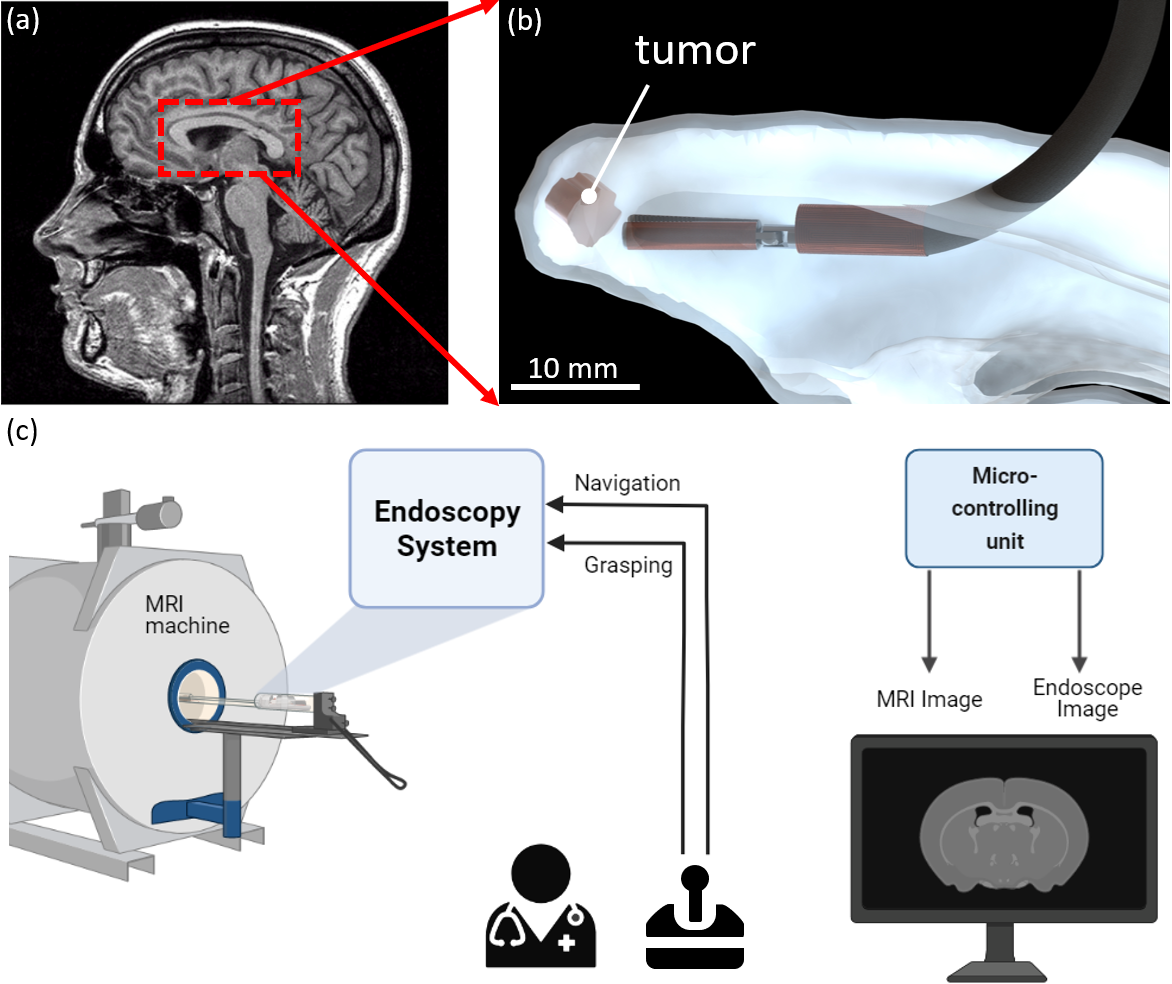}}
\caption{Overview of the MRI-driven, endoscopic system. a) MR scan locating a brain tumor within the lateral ventricle of the cerebral ventricular system. b) Computer rendering of the neuroendoscope steering towards a brain tumor using MRI-guidance. c) System overview for intraoperative MR imaging and endoscopic guidance for remote control neuroendoscopy.}
\label{Headline}
\end{figure}

Neuroendoscopy is a minimally invasive technique to visualize and treat areas of the central nervous system including the skull, brain, and spine. Currently, neuroendoscopic systems are used to treat many different kinds of cases, including intraventricular lesions, craniosynostosis, spinal lesions, and skull base tumors \cite{Shim2017}. To this end, both rigid and flexible endoscopes are utilized in surgical operations. However, the standard rigid endoscopic tools have drawbacks such as limited ﬁeld of view and risk of blunt force trauma. Neuronavigation may improve existing techniques, especially among patients with intraventricular pathologies \cite{Zimmermann2002}. Additionally, flexible endoscopes have a better mobility range which may be critical in complex anatomies like ventricles (Figure \ref{Headline}).

Although the introduction of neuroendoscopy into surgical operations has made signiﬁcant improvements over other traditionally invasive surgical approaches, there is still room for improvement, including improved precision and less trauma to surrounding tissue. Other works have integrated endoscopes to MRI systems for guidance while imaging \cite{North2012, Hill1997, Scholz1996}. The future of neuroendoscopy relies upon further miniaturization of cameras and new robotic navigation methodologies integrated with magnetic resonance intraoperative imaging to achieve better control. 

Magnetic resonance imaging (MRI) offers many benefits over other imaging approaches (fluoroscopy, ultrasound) due to its excellent soft-tissue contrast, lack of ionizing radiation, real-time tool tracking capabilities, and three-dimensional tissue information. The integration of MR-compatible surgical tools has enabled various applications in minimally invasive surgery. Studies have shown different MR-compatible device steering methods \cite{Muller2012, Yu2006} including using smart materials \cite{Sheng2018, Ho2012, Su2016}, hydraulic \cite{Fang2021, Dai2021}, pneumatic \cite{Fischer2008}, and MRI-driven (magnetic) actuation \cite{Erin2020,Tiryaki2021}. MRI-driven actuation offers significant advantages over these techniques due to its scalability, safety, response time (nearly instantaneous), accuracy (other methods experience nonlinearities in actuation), and degrees of freedom (DoF) \cite{Ali2016}. MRI-driven actuation can utilize imaging gradient coils user-controlled to generate spatial field gradients for steering a wireless robot \cite{Martel2007, Mathieu2010} or magnetic catheter tip in three dimensions (3D) \cite{Gosselin2011, Zhang2010}. Embedding magnetic elements to catheters for gradient steering introduces significant MR image distortion and additional catheter weight and bulkiness. 

The advent of MRI-driven Lorentz force-based surgical tools enabled by the magnetic field of such devices opens many opportunities for intraoperative MRI-guided neurointervention. Lorentz force-based steering approaches introduce less weight to the soft body due to the high force-to-weight ratio compared to gradient steering [55]. Moreover, the image distortion can be controlled since the image artifacts only occur when coils are activated. 
In our previous work, we demonstrated a new approach to mitigate Lorentz force-based heating effects using the inverse kinematics of the device to steer within confined workspaces \cite{Phelan2022}. This paper proposes the design and actuation of an MRI-driven endoscope for intraventricular navigation and tumor ablation. It also introduces the first Lorentz force-driven grasper for use in minimally invasive surgical interventions of the brain. Lorentz force-based steering of an endoscope under MRI guidance and direct endoscopic visualization is demonstrated within a ventricular phantom model. The endoscope is designed with a target tip orientation range with realistic medical constraints while minimizing heat dissipation to surrounding tissue.

In this paper we make the following contributions:
\begin{itemize}
  \item heat-mitigated design and development of an MRI-driven endoscope,
   \item heat-mitigated Lorentz force-based steering using integrated intraoperative MR imaging and endoscopic visualization within a human ventricular model,
   \item design and characterization of a Lorentz force-based grasper for soft tissue manipulation, and
   \item in vitro studies showing the effectiveness of Lorentz microcoils for cancerous tissue ablation.
\end{itemize}

\section{Modeling}
\subsection{MRI-driven Lorentz Force-based Actuation}
The endoscopic system designed in this study utilizes Lorentz force-based actuators for active steering and tissue grasping. The entire system can therefore be designed using the governing principles of the Lorentz force effect. Any charged particle with a given velocity in an external magnetic field experiences a force.
Using this phenomenon, a current-carrying wire experiences the Lorentz force directly proportional to its wire length L, current \textit{i}, and external magnetic field strength $\mathbf{B_0}$ as given below:

\begin{equation} \label{eq:wire_lorentz}
\mathbf F=\mathbf iL \times \mathbf B_0.
\end{equation}

In this approach, controlling microcoil current polarity directly translates to a tip deflection in the respective direction. The generated magnetic moment, $\mathbf{m}$, and corresponding torque, $\mathbf T$, can be determined in terms of the number of coil loops, \textit{N}, current, \textit{I}, and area normal vector, $\mathbf {A}$, of a coil loop, and magnetic field vector within an MR scanner ($\mathbf{B_0} = (0, 0, B_0))$, where $B_0$ is the uniform, permanent, magnetic field inside the MRI. This relation is represented as:

\begin{equation} \label{eq:lorentz_torque}
\mathbf T=\mathbf m \times \mathbf B_0 =NI (\mathbf A \times \mathbf B_0),
\end{equation}

\noindent which can be used to govern axial coil torque for a Lorentz force-actuated catheter with equally-sized coil loops. Saddle (side) microcoils can be integrated to the catheter tip for additional degrees of freedom (DoF). In this approach, both saddle coil sets are integrated on the same circumferential plane without introducing additional layer thickness compared to literature. Using the derived equation for a saddle coil's total effective area, the magnetic moment can then be expressed as:

\begin{equation} \label{eq:6}
\mathbf m = I\mathbf A \textsubscript{total}.
\end{equation}

\subsection{Cosserat Model}
\begin{figure}[t]
\centerline{\includegraphics[width=\linewidth]{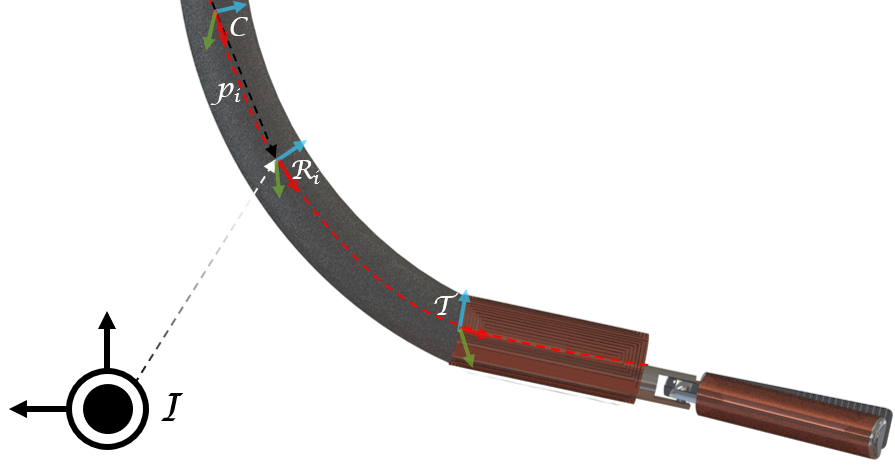}}
\caption{Computer rendering depicting the coordinate frames used in the Cosserat model for steering.}
\label{Cosserat}
\end{figure}

This paper uses the inverse kinematics of the Cosserat model to predict the tip bending angles of the proposed endoscope design as detailed in \cite{Phelan2022}. Similarly, the endoscope is modeled as a cantilever beam undergoing an external torque and tip force. We describe the state of the endoscope using a set of $N$ discretized segments $\mathbf{Y}=\left[\mathbf{y}_0^T,\mathbf{y}_1^T,...,\mathbf{y}_N^T, \right]^T_N$. The discretized state vector for each segment $i$ contains segment position ($\mathbf{p}_i\in\mathbb{R}^3$), orientation ($\mathbf{R}_i\in SO(3)$), extension force ($\mathbf{n}\in\mathbb{R}^3$), and shear torque ($\mathbf{m}\in\mathbb{R}^3$), which can be expressed in one vector  $\mathbf{y}_i:=[\mathbf{p_i},\mathbf{R_i},\mathbf{n_i},\mathbf{m_i}]$. The rotation matrix is defined in the MRI's fixed coordinate frame $\mathcal{I}$, along with two additional coordinate frames; control frame $\mathcal{C}$ representing the endoscope free length starting position and tip frame $\mathbf{T}$ locating the start of the microcoils (\textbf{Figure \ref{Cosserat}}). A system of nonlinear ordinary differential equations (ODEs) can be expressed as:  

\begin{align}  \label{eq:cosserat_equations3}
    \mathbf{\dot{p}_i} & = \mathbf{R}_i\mathbf{v}_i\\
    \mathbf{\dot{R}}_i & = \mathbf{R}_i \mathbf{u}_i\\
    \mathbf{\dot{n}}_i & = -\rho A \mathbf{g}\\
    \mathbf{\dot{m}}_i & = \mathbf{-\dot{p}}_i \times \mathbf{n}_i,
\end{align}

\noindent where $\mathbf{v}$ and $\mathbf{u}$ are tangent and curvature vectors defined as, $\mathbf{v} = \hat{\mathbf{z}} + \mathbf{K}_1\mathbf{R}^T\mathbf{n}$ and $\mathbf{u} = \mathbf{K}_2\mathbf{R}^T\mathbf{m}$, where $\hat{\mathbf{z}}$ is the unit vector in local coordinate frame, $K_1=diag(GA,GA,EA)$ and $K_2=diag(EI_A,EI_A,GJ)$. $G$, $A$, $E$, $I_A$, and $J$ represent the shear modulus, cross-sectional area, elastic modulus, area moment of inertia, and polar moment of inertia, respectively. 

\begin{align} \label{eq:torque_optim}
\argmin_{\mathbf{Y}}& ||\mathbf{R}_{\mathcal{T},des}\boxminus\mathbf{R}_\mathcal{T}||^2_2+||\tau_{des}||^2_2\\
\text{s.t.}& \mathbf{Y} = f(\mathbf{n}_0,\mathbf{m}_0). \nonumber
\end{align}

Tip torque is given as $\tau_{des}= \mathbf{m}_N$. Optimization is solved in real-time using the iterative Levenberg-Marquardt method implemented in C++ \cite{Till2019}, where endoscope forward kinematics is used as the shooting function. Endoscope forward kinematics, $\mathbf{Y} = f(\mathbf{n}_0,\mathbf{m}_0)$, can be calculated through numerical integration using a 4th order Runge-Kutta algorithm, given the endoscope's initial conditions: $\mathbf{R}_0 = R_\mathcal{C}$, $\mathbf{p}_0 =\mathbf{p}_\mathcal{C}$, $\mathbf{n}_0 = \mathbf{n}_\mathcal{C}$, and $\mathbf{m}_0 = \mathbf{m}_\mathcal{C}$.

\begin{figure*}[t]
\centerline{\includegraphics[scale=0.47]{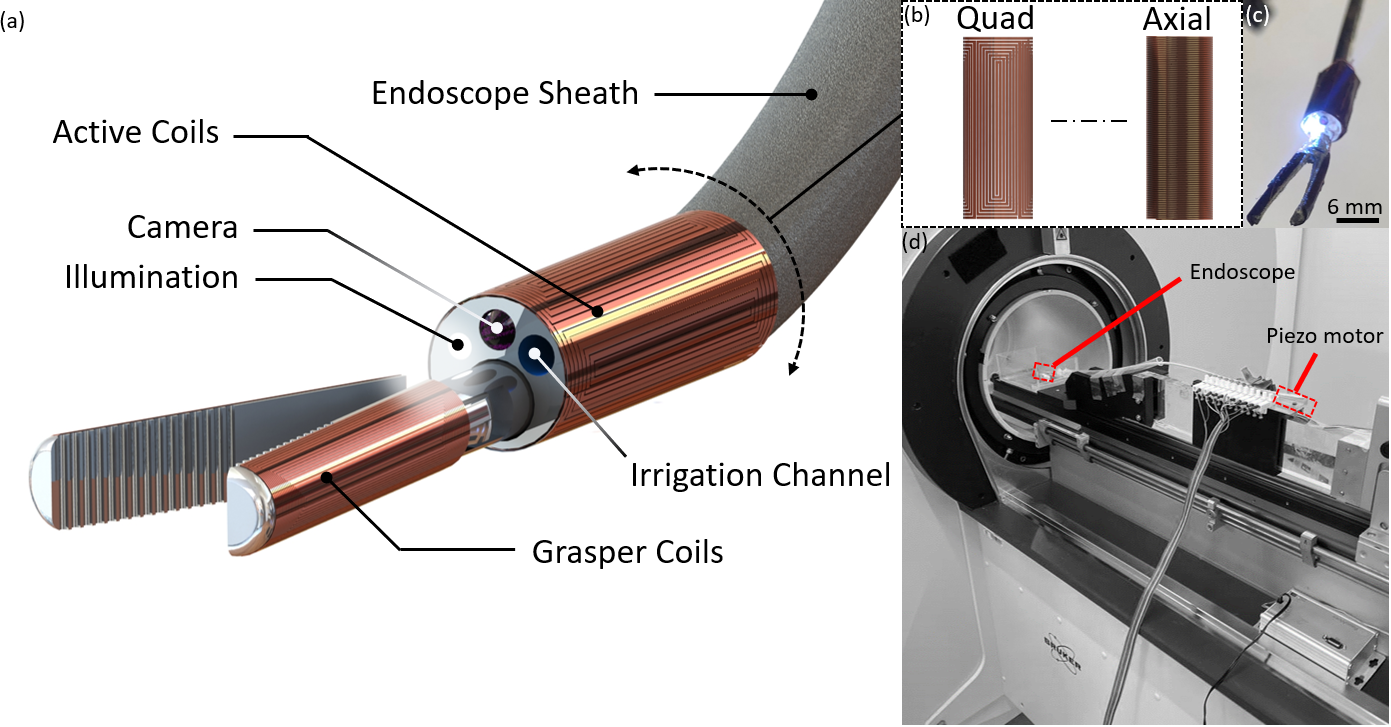}}
\caption{Endoscopic system overview. a) Computer rendering of endoscope end effector with Lorentz force-based grasping coils and active coils for remote steering. b) Assembly of active coil set consisting of one quad coil wrapped around an axial coil. c) Photograph of assembled MRI-driven endoscope. d) Photograph of MR-compatible setup with a linear piezoelectric motor to drive the active endoscope.}
\label{endoscope_photo}
\end{figure*}

\section{Methodology}
\subsection{Design Requirements}
The endoscopic system was designed to navigate the third ventricle of the brain to locate and grasp a tumor. For this application, a computer-aided rendering of the ventricular system (obtained from patient-generated MRI data) was used to create estimations of the necessary workspace. We determined a maximum tip bending angle of 90\degree\hspace{0.1em} would be sufficient to extract the tumor. The maximum endoscope length was constrained by the workspace to be less than 2 cm, which also constrains the microcoil and Lorentz grasper design. Additionally, active steering should not induce any heating effects above 1.2 W, otherwise, thermal injury could result \cite{Hetts2013}. The endoscope was also designed to enter the ventricles from a standard orientation with respect to the human brain and MRI scanner's $B_0$ field (90$\degree$). The device also needs to be MR compatible (no ferromagnetic components) therefore all integrated endoscopic components should be plastic or non-magnetic. The smallest available endoscopic camera found was the CMOS-based ScoutCam with a diameter of 1.2 mm, 160 Kpx resolution, 30 fps, 120$\degree$ field-of-view, and 5-50 mm depth of field. The endoscope tip requires irrigation to clear vision during active navigation as well as a working channel for passing tissue graspers. Lastly, there needs to be a channel for incorporating an LED.

\subsection{Endoscopic System Integration}
The MRI-driven endoscope was assembled using a Scoutcam endoscope camera attached to a stereolithography (SLA)-printed housing with an integrated irrigation channel. A white LED was attached to the endoscope cap for illumination. The Lorentz force-based grasper was assembled using specially designed microcoils manually wrapped onto SLA-printed graspers.  The steering microcoil set consisting of an axial coil bundle and a quad-coil configuration was manually wrapped around the endoscope cap to minimize the bulkiness of the endoscope tip. The entire cap assembly was secured to a 3 Fr Pellethane\textsuperscript{\textregistered} thermoplastic polyurethane tube (Nordson Medical, Salem, NH) shown in Figure \ref{endoscope_photo}.

\subsection{Microcoil Design and Fabrication}
\begin{figure}[t]
\centerline{\includegraphics[scale=0.4]{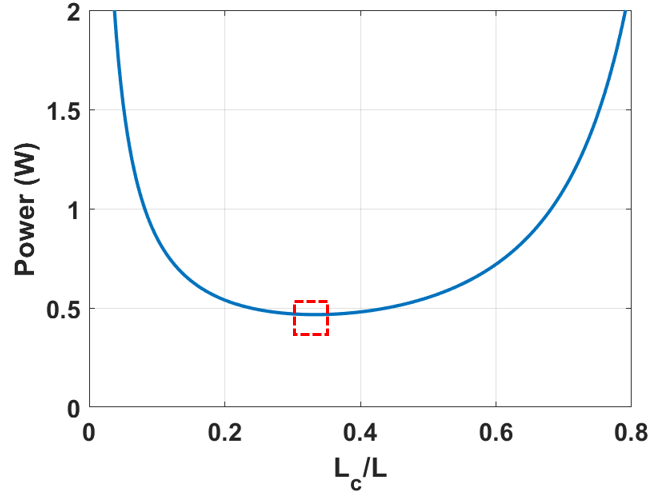}}
\caption{Design curve for determining the best suited microcoil/endoscope length to achieve 90$\degree$ of bending with minimal heat generation. The red box indicates the optimal ratio of 0.33.}
\label{Coil_length}
\end{figure}

The active microcoils integrated to the endoscope cap were optimized according to the power optimization algorithm found in our previous work, however, as previously mentioned, the maximum desired tip orientation would be 90$\degree$ for this application.  With this goal in mind, a coil length design curve was generated to find the optimal coil/endoscope ratio for minimizing heat generation during steering. A value of 0.33 was determined (Figure \ref{Coil_length}) but to ease manufacturing, a ratio of 0.5 was used in this study with additional parameters found in Table \ref{cathparams}.
Microcoil bundles were manufactured starting with the axial coil forming the outer core of the endoscope cap, following the same fabrication process as previously described \cite{Phelan2022}. The resulting coil bundle was attached to the endoscope cap as shown in Figure \ref{endoscope_photo}.  The final prototype was a 3-DoF tri-coil endoscope with a final outer diameter of approximately 4.5 mm (Figure \ref{endoscope_photo}(c)).

\subsection{Lorentz Force-Based Grasper Design}
\begin{figure*}[t]
\centerline{\includegraphics[width=\linewidth]{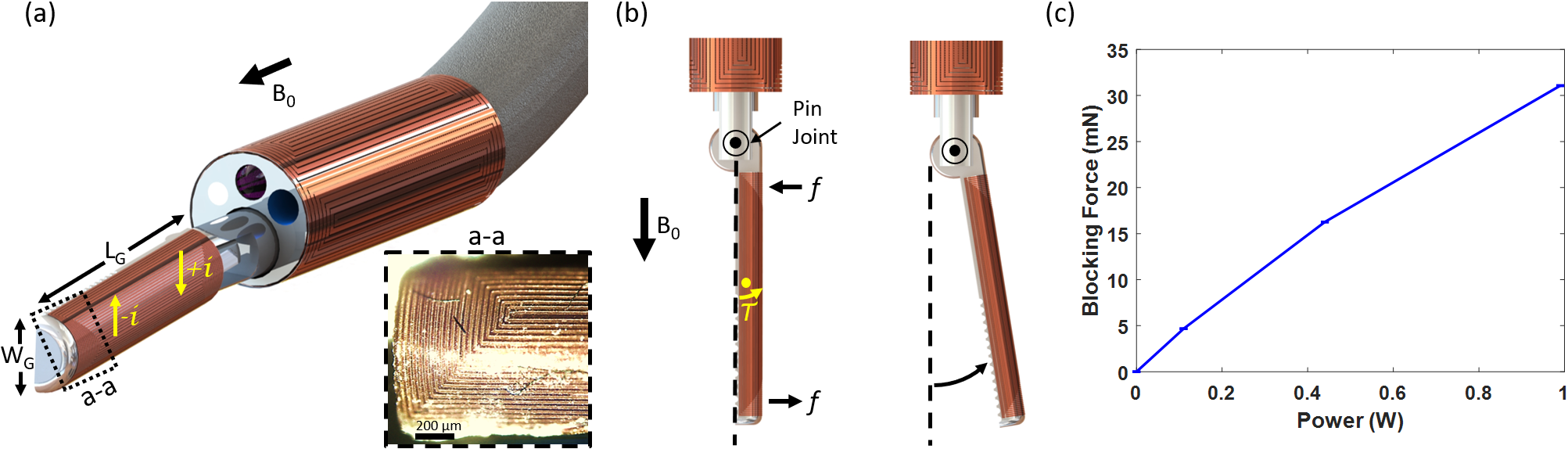}}
\caption{Lorentz force-based grasping model. a) Upon an induced current, copper wire perpendicular to the MRI $B_0$ field field experiences the Lorentz force in equal and opposite directions. These forces cause a net torque on the body as shown in (b), proportional to the length, $L_G$, and width, $W_G$, of the coils. a-a) A photograph of the laser-cut grasper microcoil. c) Blocking force measurements obtained from applying a range of power inputs to the Lorentz grasper.}
\label{Lorentz_gripper}
\end{figure*}

Other studies have shown the range of forces necessary for grasping tissue from dissecting nervous tissue (0.4 N) to connective tissue (45.8 N) \cite{Golahmadi2021}.  However, the maximum force generated by the Lorentz grasper, or blocking force, is unknown. The blocking force was determined experimentally using a non-magnetic force sensor (LSB200, Futek) within the MRI. To maximize this value, Equation \ref{eq:6} dictates that the net torque induced by the microcoil is directly proportional to the area, current, and number of coil turns. Therefore, a grasping coil was specially designed (Figure \ref{Lorentz_gripper}) to maximize magnetic moment using the parameters given in Table \ref{cathparams}. The grasper was then attached on a pin joint and secured in place next to the force sensor. Power inputs up to 1 W were induced for three separate trials in air and the resultant blocking force was measured. 



\subsection{MR Imaging}
All MRI imaging experiments were done using a 7T (Bruker, Ettlingen, Germany), actively shielded gradient (BGA20SHP) preclinical system with a maximum 300 mT/m gradient strength and 154 mm quadrature birdcage coil. For the phantom model scan and maximum intensity projection reconstruction, 3D rapid fast spin echo sequence was used (TR/TEeff 1500/68.93 ms, NEX = 1, RARE factor = 16, 0.512 x 0.61 x 0.65 mm$^3$ spatial resolution, 21 min scan duration).  For endoscopic steering demos, a fast spin echo sequence (TR/TEeff 500/37.03 ms, RARE factor = 64, NEX = 1, 0.70 x 0.55 mm$^2$ spatial resolution, 2 mm slice thickness, 1 slice, total duration: 500 ms) was used.  

\subsection{Cancerous Tissue and Ventricular Phantom Preparation}
Human SH-SY5Y neuroblastoma cells (American Type Culture Collection) were cultured in Dulbecco's Modified Eagle's Medium (DMEM) supplemented with 10$\%$ FBS (Gibco) and 0.5$\%$ penicillin/streptomycin (Gibco) in a humidified, 37°C and 5$\%$ $CO_2$ environment. Cells were used at passage numbers lower than 10, and surface detachment during splitting was performed using 0.05$\%$ trypsin-EDTA (Gibco) when 80$\%$ confluency was reached.

For the preparation of tumor spheroids, Nunclon Sphera-treated, U-shaped-bottom 96-well microplates (Thermo Scientific) were used. SH-SY5Y cells were seeded into spheroid microplates at densities of 150,000 cells/well. The cells were grown for 48 h in a humidified, 37°C and 5$\%$ $CO_2$ environment. The SH-SY5Y spheroids' diameter was measured at approximately 1 mm after 48 h. The prepared tumor spheroids were ﬁnally injected into the SLA-printed phantom towards the lateral ventricle of the brain and sealed off with blue-colored water to ensure the phantom was fully submerged during imaging.

\subsection{Tumor Ablation}
The objective of this study was to determine the minimum power input needed to ablate cancerous tissue for future experiments. Therefore, initial heat ablation experiments were conducted in a fume hood, in which tumor spheroids were treated with different power values (0.03 mW, 0.11 mW, 0.44 mW and 0.69 mW) for three  trials each lasting 1 min. The amount of time necessary for ablation can vary based on different factors including the electrodes' diameter, tumor size, and power input. However, literature shows average times between 4 and 6 minutes and some even longer \cite{Rivas2021}, while clinical trials have demonstrated ablation times below 4 minutes.
The power values were calculated using the microcoil's resistance (11 $\Omega$) and corresponding current values (50 mA, 100 mA, 200 mA, and 0.25 mA).  After heating, tumor spheroids were incubated again at 37°C for the following 24 h. Next, the viabilities of tumor spheroids were measured using a CellTiter-Glo 3D Cell Viability Assay. A volume of CellTiter-Glo 3D Reagent equal to the volume of cell culture medium was added to the test groups and mixed for 5 min to perform cell lysis. The plate was incubated at room temperature for an additional 25 min to stabilize the luminescent signal. The luminescence values were measured in an opaque 96-well plate using a plate reader (BioTek's Synergy 2, Winooski, VT, USA). The cell viability percentages were calculated based on the untreated group (cell viability = 100$\%$). 

\renewcommand{\arraystretch}{1.2}
\begin{table}
\caption{Endoscope Design Parameters ($\theta_{des}$ = 90\degree)}
\label{cathparams}
\begin{center}
\begin{tabular}{c c c}
\hline
\hline
Parameter & Description & Value\\
$W_c$ & Saddle coil width & 0.76 mm\\
$D_c$ & Endoscope diameter & 3.5 mm\\
$t$ & Wire gap spacing & 40 $\mu$m\\
$C_{cw}$ & Coil core width & 150 $\mu$m\\
$s$ & Wire thickness & 18 $\mu$m\\
$w$ & Wire width & 40 $\mu$m\\
$N_{saddle}$ & Saddle coil turns & 7\\
${pw}_{D}$ & Power wire diameter  & 80 $\mu$m\\
\hline
$L_c$ & Coil length & 10 mm\\
$N_{axial}$ & Axial coil turns & 250\\
$I_{saddle}$ & Saddle coil current & 0 mA\\
$I_{axial}$ & Axial coil current & 213 mA\\
$P_{total}$ & Total power & 0.4663 W\\
\hline
\hline
$W_G$ & Grasper coil width & 3.1 mm\\
$t_G$ & Wire gap spacing & 15 $\mu$m\\
$G_{cw}$ & Grasper coil core width & 100 $\mu$m\\
$L_G$ & Grasper coil length & 10 mm\\
$N_{G}$ & Grasper coil turns & 20\\
\hline
\hline
\end{tabular}
\end{center}
\end{table}


\section{Experimental Results}
\subsection{Workspace Characterization and Camera Steering}
The workspace of the active endoscope was characterized using an MR-compatible camera (12M, MRC) from the top view. The endoscope was controlled using two dual motor drivers (DRV8835, Pololu) for active steering and one driver for grasping. Linear actuation was provided by a nonmagnetic piezoelectric motor (LS15, Piezomotor) (Figure \ref{endoscope_photo}(d)). The endoscope was fixed to the linearly sliding, manual bed of the MRI system. All power cables were extended from the MRI to the adjacent computer room. The endoscope was constrained at an insertion length of 3 cm and suspended in a stagnant water bath. Current values up to 300 mA were tested which is sufficient for the purposes of this study to validate the experimental flexural rigidity. This value was determined to be 4.45e-5 Nm$^2$ and was used to inform the inverse kinematic controller for reaching angles up to 90\degree.

\begin{figure*}[t]
\centerline{\includegraphics[scale=0.3]{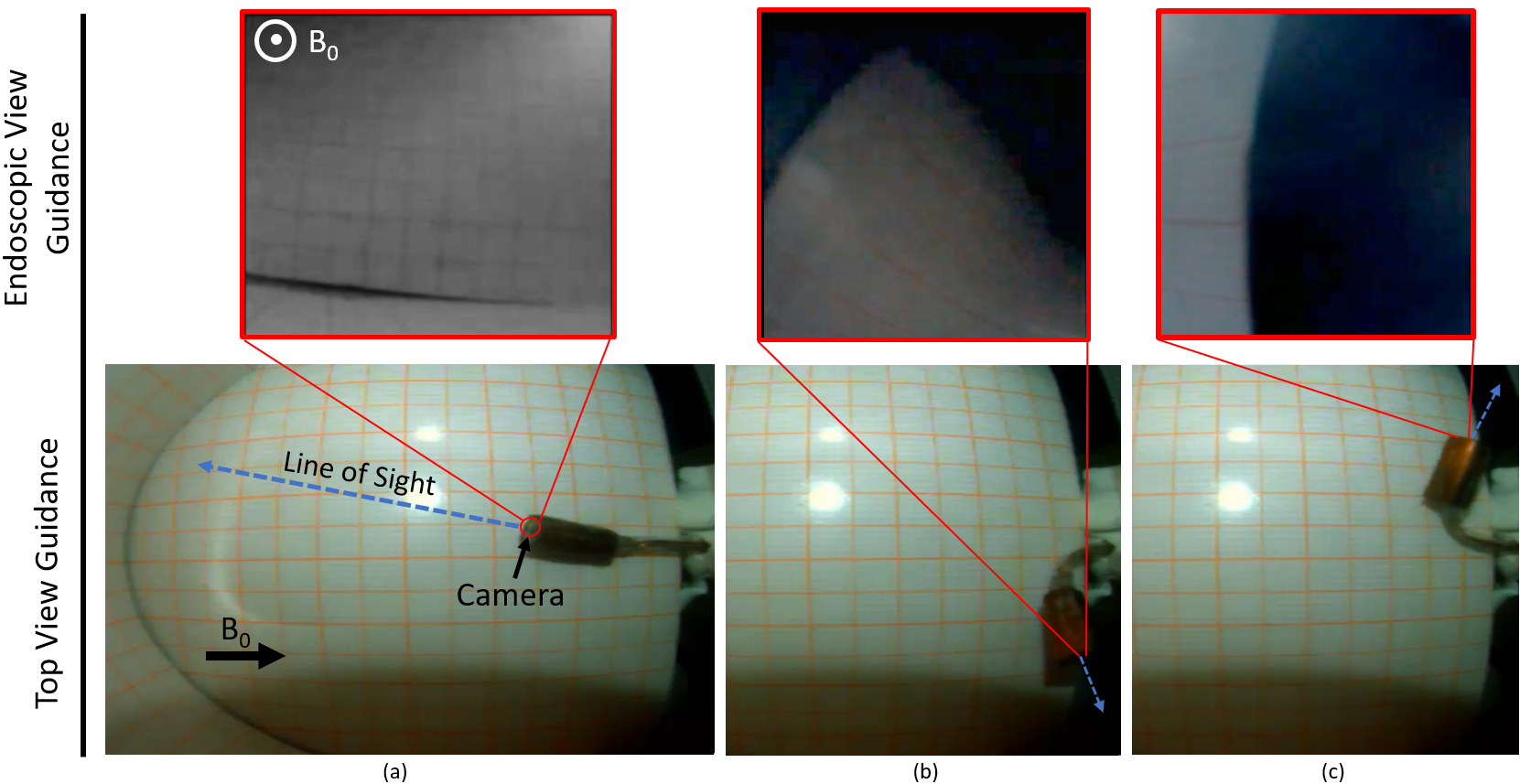}}
\caption{Camera steering of the endoscope using the power-optimized controller with endoscopic and top view perspectives. a) The endoscope begins aligned with the $B_0$ field with the camera pointing towards the grid pattern in front. The endoscope was steered to the left (b) and right (c) using axial and side coils collectively.}
\label{camera_steering}
\end{figure*}

To demonstrate camera steering, the endoscope was steered left and right to achieve 100\degree\hspace{0.1em} tip orientations as shown in Figure \ref{camera_steering}. To achieve such orientations, the axial and side coils were used in combination to move through the workspace. Negative currents were induced to the axial coil while the polarity of the side coil has to be changed upon reaching 90\degree\hspace{0.1em} for steering in both directions. The power-optimized current controller regulates axial and side coil usage respectively to minimize heat dissipation. In addition, the endoscopic viewpoint can be used to monitor vertical motions  since the endoscope may not be perfectly aligned during its initial orientation.

\begin{figure*}[t]
\centerline{\includegraphics[scale=0.28]{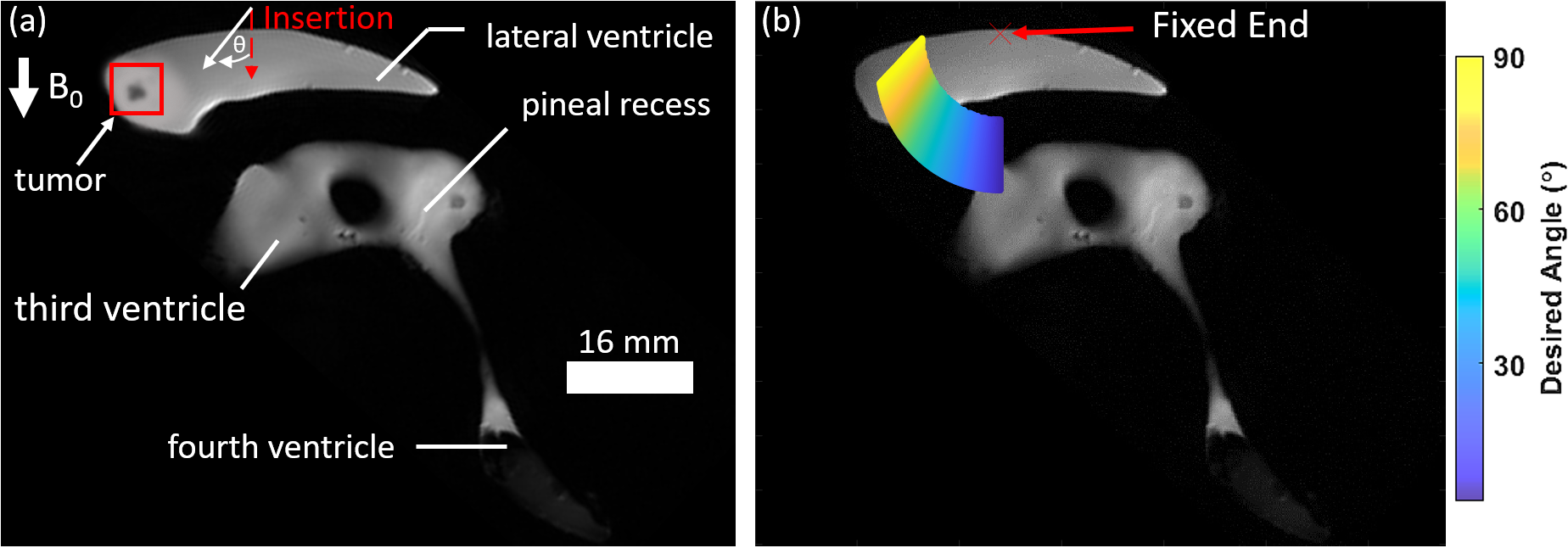}}
\caption{Figures showing the intraventricular workspace of the neuroendoscope. a) Preoperative MRI scan of the phantom model displaying an overview of the ventricular system with the red line depicting the entry angle of the endoscope with respect to the MRI $B_0$ field. b) Theoretical workspace of the active endoscope using proposed model overlaid on MRI scan. The model predicts the endoscope can reach the desired location at the maximum angle displayed in yellow.}
\label{workspace}
\end{figure*}

\begin{figure*}[t]
\centerline{\includegraphics[scale=0.4]{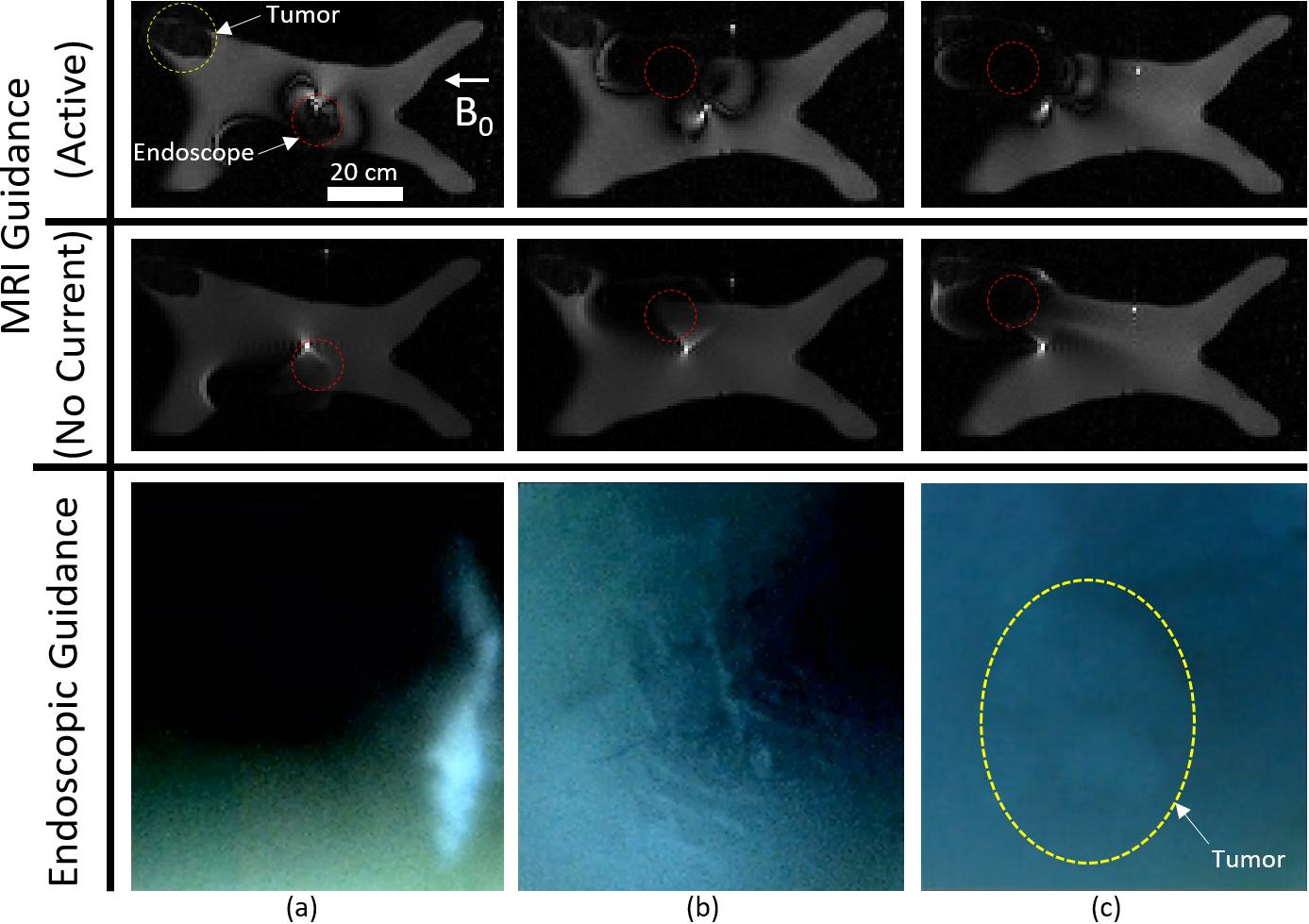}}
\caption{MRI-guided active imaging through the ventricular system. a) MR scan showing the lateral ventricle and tumor (highlighted in yellow) and the entry point of the endoscope (in red). The tip of the endoscope can be tracked when the coils are actively being steered using the distortion pattern in the image. This pattern can be controlled by turning coils off during imaging. The endoscope first enters the lateral ventricle and begins steering towards the tumor (b). c) The tumor was located using the MR scan and visibly observed and treated using direct visualization from the endoscope camera.}
\label{active_imaging}
\end{figure*}

\subsection{Lorentz Grasping Force Characterization}
 The blocking force generated by the Lorentz grasper was characterized as shown in Figure \ref{Lorentz_gripper}(c). Current inputs up to 0.5 A were induced to the Lorentz grasper and measured in mN. Results indicated a maximum measurable force exertion of 31 mN due to excessive heating of the grasper, however, forces below 20 mN are more than sufficient for making soft tissue contact. Power inputs above this threshold (0.5 W) can be used for thermal ablation of diseased tissue.

\subsection{Tumor Spheroid Ablation}
We performed heat ablation of tumor spheroids using electrical current inputs. The tumor spheroids were exposed to the endoscopic grasping microcoils, and Joule heating resulted in thermal ablation. The 3D cytotoxicity analyses of the tumor spheroids indicated that we could decrease the cellular viabilities to 16$\%$ with a power input of 0.69 mW using the proposed neuroendoscope (Figure 9).

\subsection{MRI-Guided Intraventricular Navigation}
In order to navigate the ventricular phantom, a preoperative scan of the ventricular phantom locating the brain tumor was taken (Figure \ref{workspace}(a)) and the workspace was simulated for the device to reach the target (Figure \ref{workspace}(b)). The neuroendoscope was steered using joystick control while MRI and endoscopic guidance is provided simultaneously (Figure \ref{active_imaging}). The endoscope was inserted through the lateral ventricle to replicate an intraventricular neuroendoscopic procedure. Then the power-optimized controller assisted the user in controlling the orientation of the endoscope to navigate towards the tumor. The endoscope was tracked using the image distortion pattern induced by active coils during steering. To obtain an image without distortion and observe the environment, the steering coils were powered off. Results and video feedback show that the endoscope was successfully guided using MR imaging and direct visualization to locate the brain tumor. The endoscopic graspers would then be used in the future to grasp and thermally ablate the tumor using the previously determined 0.69 mW.

\begin{figure}
\centerline{\includegraphics[scale=0.4]{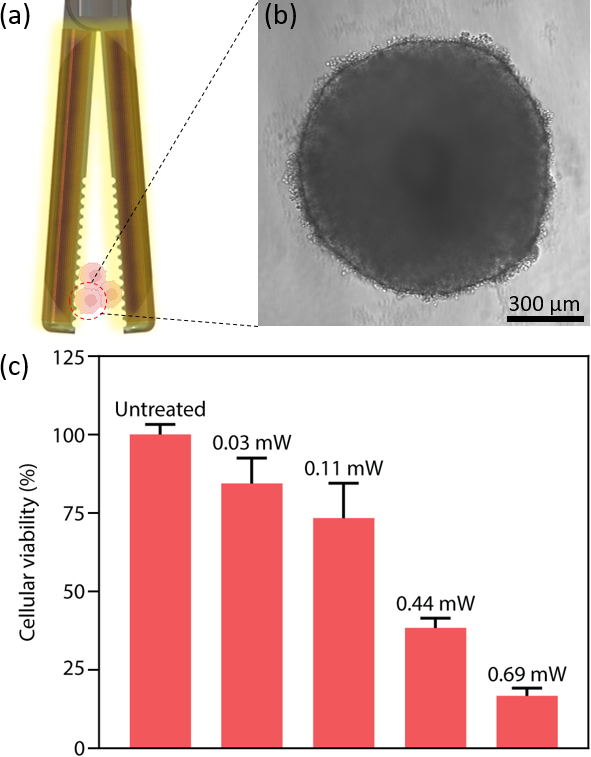}}
\caption{Lorentz force-based Joule heating for tumor ablation. a) The Joule heating effect from grasping coils can be controlled to ablate cancerous tissue upon contact. b) Microscope photograph showing a representative tumor spheroid. c) Tumor cell viability test showing the effectiveness of Joule heating-induced ablation.}
\label{tumor_ablation}
\end{figure}

\section{Discussion}

This work has demonstrated the capabilities of an integrated robotic system for MRI-guided neuroendoscopy. The endoscope was characterized and controlled using the inverse kinematic controller and demonstrated maneuverability up to 90\degree\hspace{0.1em} given power safety limits. However, an irrigation port has been integrated into the working channel of the endoscope, enabling active cooling of the endoscopic tip. This would allow for increased degrees of bending depending upon anatomical constraints. Although the endoscope's final size is comparable to existing commercial devices, further improvements can be made such as incorporating a fiber optic for illumination, using photolithography methods to pattern the microcoils directly on the cap, and fabricating a flexible cap to make the design more versatile for other applications. Additionally, Joule-heating could be mitigated further by increasing the copper thickness of the active coils, implementation of an active steering method with real-time temperature sensing with active cooling to regulate device temperature, or system integration using MR thermometry for real-time tissue temperature monitoring. 

A Lorentz force-based grasper was designed and proved to be effective for making contact and ablating cancerous tissue. Blocking force measurements demonstrated the device can provide reasonable tissue contact forces up to 31 mN, which is comparable to forces required for sharp dissection and clamping, but further improvements can be made to maximize force output for vascular/neurosurgery (up to 1 N) \cite{Golahmadi2021}. Higher forces were not demonstrated due to Joule heating effects in air causing wire detachment, but cooling in aqueous environments like the body can permit higher grasping forces. Furthermore, orientation of the grasper with respect to the endoscope and MRI $B_0$ field should also be considered and requires further study. A similar triaxial coil design for grasping could maximize such force capabilities as demonstrated for active steering. The Lorentz grasping microcoils were also effective at ablating tumor tissue at 16$\%$ cell viability with low power (0.69 mW). This value is widely acceptable for 3D cell cultures \cite{Baek2016, Landgraf2022}. Moreover, spheroid cell viability could be decreased by applying current or measuring viability over a longer period of time. These results suggest the potential of diseased tissue ablation while grasping for improved treatments. It also suggests the possibility of cauterizing open wounds for preventing blood loss during procedures.

The neuroendoscope was steered remotely under MRI and endoscopic guidance simultaneously to survey the workspace of the ventricular system. Using MRI guidance, a tumor was located within the lateral ventricle and navigated in real-time. Upon reaching the diseased tissue area, endoscopic visual feedback confirmed the tumor had been reached and treatment could proceed, including grasping and ablation. Further improvements in endoscopic vision due to limited contrast from the endoscope would allow for more precise navigation and detection methods. This study has proven integrated intraoperative imaging in conjunction with real-time image processing is feasible for improved surgical accuracy and enhanced treatment of diseased tissue. 




\section{Conclusion}
Neuroendoscopy can benefit from improved steerable devices for treating tissue pathologies within the central nervous system. This paper described the design and development of an MRI-driven endoscope for neuroendoscopy. It has been demonstrated that the power-optimized controller can be extended to endoscopic operations within constrained workspaces found in the ventricular system of the brain. The feasibility of a Lorentz force-based grasper was also evaluated and shown to be effective at making contact and ablating cancerous tissue. These Lorentz force-based design and actuation methodologies further extend the capabilities of MRI-driven robotics for integrated intraoperative MR imaging and endoscopy.

\section*{Acknowledgments}
The authors thank A. Aydin for discussions on neuroendoscopy and M. Rolon for assistance with manuscript preparation.

\bibliographystyle{IEEEtran} 
\bibliography{endoscope} 

\begin{thebibliography}{10}
\providecommand{\url}[1]{#1}
\csname url@rmstyle\endcsname
\providecommand{\newblock}{\relax}
\providecommand{\bibinfo}[2]{#2}
\providecommand\BIBentrySTDinterwordspacing{\spaceskip=0pt\relax}
\providecommand\BIBentryALTinterwordstretchfactor{4}
\providecommand\BIBentryALTinterwordspacing{\spaceskip=\fontdimen2\font plus
\BIBentryALTinterwordstretchfactor\fontdimen3\font minus
  \fontdimen4\font\relax}
\providecommand\BIBforeignlanguage[2]{{%
\expandafter\ifx\csname l@#1\endcsname\relax
\typeout{** WARNING: IEEEtran.bst: No hyphenation pattern has been}%
\typeout{** loaded for the language `#1'. Using the pattern for}%
\typeout{** the default language instead.}%
\else
\language=\csname l@#1\endcsname
\fi
#2}}

\bibitem{Shim2017}
K.~W. Shim, E.~K. Park, D.~S. Kim, and J.~U. Choi, ``{Neuroendoscopy: Current
  and future perspectives},'' \emph{Journal of Korean Neurosurgical Society},
  vol.~60, no.~3, pp. 322--326, 2017.

\bibitem{Zimmermann2002}
M.~Zimmermann, D.~Ph, A.~Raabe, and D.~Ph, ``{Robot-assisted Navigated
  Neuroendoscopy},'' vol.~51, no.~6, pp. 1446--1452, 2002.

\bibitem{North2012}
O.~J. North, M.~Ristic, C.~A. Wadsworth, I.~R. Young, and S.~D.
  Taylor-Robinson, ``{Design and evaluation of endoscope remote actuator for
  MRI-guided Endoscopic Retrograde Cholangio-Pancreatography (ERCP)},''
  \emph{Proceedings of the IEEE RAS and EMBS International Conference on
  Biomedical Robotics and Biomechatronics}, pp. 787--792, 2012.

\bibitem{Hill1997}
D.~L.~G. Hill, L.~A. Langsaeter, P.~N. Poynter-Smith, C.~L. Emery, P.~E.
  Summers, S.~F. Keevil, J.~P.~M. Pracy, R.~Walsh, D.~J. Hawkes, and M.~J.
  Gleeson, ``{Feasibility Study of Magnetic Resonance Imaging-Guided Intranasal
  Flexible Microendoscopy},'' \emph{Computer Aided Surgery}, vol.~2, no.~5, pp.
  264--275, 1997.

\bibitem{Scholz1996}
M.~Scholz, M.~Deli, U.~Wildf{\"{o}}rster, K.~Wentz, A.~Recknagel,
  H.~Preuschoft, and A.~Harders, ``{MRI-guided endoscopy in the brain: A
  feasability study},'' \emph{Minimally Invasive Neurosurgery}, vol.~39, no.~2,
  pp. 33--37, 1996.

\bibitem{Muller2012}
L.~Muller, M.~Saeed, M.~W. Wilson, and S.~W. Hetts, ``{Remote control catheter
  navigation: Options for guidance under MRI},'' \emph{Journal of
  Cardiovascular Magnetic Resonance}, vol.~14, no.~1, pp. 1--9, 2012.

\bibitem{Yu2006}
N.~Yu and R.~Riener, ``{Review on MR-compatible robotic systems},''
  \emph{Proceedings of the First IEEE/RAS-EMBS International Conference on
  Biomedical Robotics and Biomechatronics, 2006, BioRob 2006}, vol. 2006, pp.
  661--665, 2006.

\bibitem{Sheng2018}
J.~Sheng, X.~Wang, T.-M.~L. Dickfeld, and J.~P. Desai, ``{Towards the
  Development of a Steerable and MRI-Compatible Cardiac Catheter for Atrial
  Fibrillation Treatment},'' \emph{IEEE Robotics and Automation Letters},
  vol.~3, no.~4, pp. 4038--4045, 2018.

\bibitem{Ho2012}
M.~Ho, A.~B. McMillan, J.~M. Simard, R.~Gullapalli, and J.~P. Desai, ``{Toward
  a meso-scale SMA-actuated MRI-compatible neurosurgical robot},'' \emph{IEEE
  Transactions on Robotics}, vol.~28, no.~1, pp. 213--222, 2012.

\bibitem{Su2016}
H.~Su, G.~Li, D.~C. Rucker, R.~J. Webster, and G.~S. Fischer, ``{A Concentric
  Tube Continuum Robot with Piezoelectric Actuation for MRI-Guided Closed-Loop
  Targeting},'' \emph{Annals of Biomedical Engineering}, vol.~44, no.~10, pp.
  2863--2873, 2016.

\bibitem{Fang2021}
G.~Fang, M.~C. Chow, J.~D. Ho, Z.~He, K.~Wang, T.~C. Ng, J.~K. Tsoi, P.~L.
  Chan, H.~C. Chang, D.~T.~M. Chan, Y.~H. Liu, F.~C. Holsinger, J.~Y.~K. Chan,
  and K.~W. Kwok, ``{Soft robotic manipulator for intraoperative MRI-guided
  transoral laser microsurgery},'' \emph{Science Robotics}, vol.~6, no.~57,
  2021.

\bibitem{Dai2021}
J.~Dai, Z.~He, G.~Fang, X.~Wang, Y.~Li, C.~L. Cheung, L.~Liang, I.~Iordachita,
  H.~C. Chang, and K.~W. Kwok, ``{A Robotic Platform to Navigate MRI-guided
  Focused Ultrasound System},'' \emph{IEEE Robotics and Automation Letters},
  vol.~6, no.~3, pp. 5137--5144, 2021.

\bibitem{Fischer2008}
G.~S. Fischer, I.~Iordachita, C.~Csoma, J.~Tokuda, S.~P. DiMaio, C.~M. Tempany,
  N.~Hata, and G.~Fichtinger, ``{MRI-compatible pneumatic robot for
  transperineal prostate needle placement},'' \emph{IEEE/ASME Transactions on
  Mechatronics}, vol.~13, no.~3, pp. 295--305, 2008.

\bibitem{Erin2020}
O.~Erin, M.~Boyvat, M.~E. Tiryaki, M.~Phelan, and M.~Sitti, ``{Magnetic
  Resonance Imaging System-Driven Medical Robotics},'' \emph{Advanced
  Intelligent Systems}, vol.~2, p. 1900110, 2020.

\bibitem{Tiryaki2021}
M.~E. Tiryaki and M.~Sitti, ``{Magnetic Resonance Imaging‐Based Tracking and
  Navigation of Submillimeter‐Scale Wireless Magnetic Robots},''
  \emph{Advanced Intelligent Systems}, p. 2100178, 2021.

\bibitem{Ali2016}
A.~Ali, D.~H. Plettenburg, and P.~Breedveld, ``{Steerable Catheters in
  Cardiology: Classifying Steerability and Assessing Future Challenges},''
  \emph{IEEE Transactions on Biomedical Engineering}, vol.~63, no.~4, pp.
  679--693, 2016.

\bibitem{Martel2007}
S.~Martel, J.~B. Mathieu, O.~Felfoul, A.~Chanu, E.~Aboussouan, S.~Tamaz,
  P.~Pouponneau, L.~Yahia, G.~Beaudoin, G.~Soulez, and M.~Mankiewicz,
  ``{Automatic navigation of an untethered device in the artery of a living
  animal using a conventional clinical magnetic resonance imaging system},''
  \emph{Applied Physics Letters}, vol.~90, no.~11, pp. 11--13, 2007.

\bibitem{Mathieu2010}
J.~B. Mathieu and S.~Martel, ``{Steering of aggregating magnetic microparticles
  using propulsion gradients coils in an MRI scanner},'' \emph{Magnetic
  Resonance in Medicine}, vol.~63, no.~5, pp. 1336--1345, 2010.

\bibitem{Gosselin2011}
F.~Gosselin, V.~Lalande, and S.~Martel, ``{Characterization of the deflections
  of a catheter steered using a magnetic resonance imaging system},''
  \emph{Medical Physics}, vol.~38, no.~9, pp. 4994--5002, 2011.

\bibitem{Zhang2010}
K.~Zhang, A.~J. Krafft, R.~Umathum, F.~Maier, W.~Semmler, and M.~Bock,
  ``{Real-time MR navigation and localization of an intravascular catheter with
  ferromagnetic components},'' \emph{Magnetic Resonance Materials in Physics,
  Biology and Medicine}, vol.~23, pp. 153--163, 2010.

\bibitem{Phelan2022}
M.~F. Phelan, M.~E. Tiryaki, J.~Lazovic, H.~Gilbert, and M.~Sitti,
  ``{Heat‐Mitigated Design and Lorentz Force‐Based Steering of an
  MRI‐Driven Microcatheter toward Minimally Invasive Surgery},''
  \emph{Advanced Science}, vol.~9, no.~10, p. 2105352, 2022.

\bibitem{Till2019}
J.~Till, V.~Aloi, and C.~Rucker, ``{Real-time dynamics of soft and continuum
  robots based on Cosserat rod models},'' \emph{International Journal of
  Robotics Research}, vol.~38, pp. 723--746, 2019.

\bibitem{Hetts2013}
S.~W. Hetts, M.~Saeed, A.~J. Martin, L.~Evans, A.~F. Bernhardt, V.~Malba,
  F.~Settecase, L.~Do, E.~J. Yee, A.~Losey, R.~Sincic, P.~Lillaney, S.~Roy,
  R.~L. Arenson, and M.~W. Wilson, ``{Endovascular catheter for magnetic
  navigation under MR imaging guidance: Evaluation of safety in vivo at
  1.5T},'' \emph{American Journal of Neuroradiology}, vol.~34, no.~11, pp.
  2083--2091, 2013.

\bibitem{Golahmadi2021}
A.~K. Golahmadi, D.~Z. Khan, G.~P. Mylonas, and H.~J. Marcus, ``{Tool-tissue
  forces in surgery: A systematic review},'' \emph{Annals of Medicine and
  Surgery}, vol.~65, p. 102268, 2021.

\bibitem{Rivas2021}
R.~Rivas, R.~B. Hijlkema, L.~J. Cornelissen, T.~C. Kwee, P.~C. Jutte, and P.~M.
  van Ooijen, ``{Effects of control temperature, ablation time, and background
  tissue in radiofrequency ablation of osteoid osteoma: A computer modeling
  study},'' \emph{International Journal for Numerical Methods in Biomedical
  Engineering}, vol.~37, no.~9, pp. 1--14, 2021.

\bibitem{Baek2016}
N.~H. Baek, O.~W. Seo, M.~S. Kim, J.~Hulme, and S.~S.~A. An, ``{Monitoring the
  effects of doxorubicin on 3D-spheroid tumor cells in real-time},''
  \emph{OncoTargets and Therapy}, vol.~9, pp. 7207--7218, 2016.

\bibitem{Landgraf2022}
L.~Landgraf, A.~Kozlowski, X.~Zhang, M.~Fournelle, F.~J. Becker, S.~Tretbar,
  and A.~Melzer, ``{Focused Ultrasound Treatment of a Spheroid In Vitro Tumour
  Model},'' \emph{Cells}, vol.~11, no.~9, pp. 1--14, 2022.

\end{thebibliography}

\end{document}